\documentclass[pra,twocolumn,final,nototal,superscriptaddress,amsmath,amsfonts,noindent]{revtex4}
\usepackage{graphicx}

\newcommand{\figbox}[1]{%
  \fbox{%
    \vbox to 1in{%
    \vfil
    \hbox to 2in{%
      \hfil
      #1%
      \hfil}%
    \vfil}}}

\begin{document}

\title{Feshbach molecules in a one-dimensional Fermi gas}

\author{D.B.M. Dickerscheid}
\affiliation{Institute for Theoretical Physics, University of Utrecht,
Princetonplein 5, 3584 CC Utrecht, The Netherlands} 
\email{dickrsch@phys.uu.nl}
\author{H. T. C. Stoof}
\affiliation{Institute for Theoretical Physics, University of Utrecht,
Princetonplein 5, 3584 CC Utrecht, The Netherlands} 

\begin{abstract}
We consider the binding energy and the wave function of Feshbach molecules confined in a one-dimensional matter 
waveguide. We compare the binding energy with the experiment of Moritz \emph{et al.} \cite{Moritz2005a} and find
excellent agreement for the full magnetic field range explored 
experimentally.
\end{abstract}

\maketitle 

\section{Introduction}

In a beautiful experiment Moritz \emph{et al.} recently
reported the observation of two-particle bound 
states of  ~$^{40}$K confined in a one-dimensional
matter waveguide  \cite{Moritz2005a}. 
In the experiment an array of equivalent one-dimensional quantum system 
is realized by trapping a mixture of two hyperfine states of  
 ~$^{40}$K atoms in a two-dimensional optical lattice.
The atoms are trapped at the intensity maxima and the radial confinement 
is only a fraction of the lattice period.
At a given value of the magnetic field the binding energy $E_{B}$ of 
the bound states is probed by radio-frequency spectroscopy.

Although Moritz \emph{et al.}  realized its limitations,
the description of the experiment makes use of 
a single-channel model of radially confined atoms interacting 
with a pseudopotential \cite{Wilkens, Bergeman2003a}. Within this model
the bound-state energy $E_{B}$ is related to the s-wave scattering length $a$ of the atoms
by \begin{eqnarray}\label{eq1}
\frac{a}{a_{\perp}} = - \frac{\sqrt{2}}{\zeta(1/2, 1/2 -E_{B}/2 \hbar \omega_{\perp})},
\end{eqnarray}
where $a_{\perp} = \sqrt{\hbar/m \omega_{\perp}}$,
$m$ is the atomic mass, and $\omega_{\perp}$ is the radial 
trapping frequency.
To vary the scattering length, however, the experiment makes use of
a Feshbach resonance at a magnetic field of $B_{0} = 202.1$~Gauss.
For such a Feshbach resonance a two-channel approach is physically
more realistic.

For the Feshbach problem the molecular binding energy $E_{B}$ always satisfies
the equation \cite{review},
\begin{equation}
E_{B} - \delta (B)  = \hbar \Sigma (E_{B}).
\end{equation}
Here the detuning $\delta(B) = \Delta \mu (B - B_{0})$ 
varies as a function of the magnetic field and depends on the 
difference in magnetic moments $\Delta \mu$ 
between the open and closed channels in the Feshbach problem.
The resonance is located at the magnetic field strength $B_{0}$.
For the homogeneous Fermi gas 
the molecular selfenergy is given by \cite{review}
\begin{eqnarray}
\hbar \Sigma (E) = - \left( \frac{g^{2} m^{3/2} }{ 4 \pi \hbar^{3}} \right) 
\frac{i \sqrt{E} }{ 1 - i |a_{\rm bg}| \sqrt{m E/\hbar^{2}} },
\end{eqnarray}
which leads to corrections to the single-channel result 
$-\hbar^{2} / m a^{2}$.  
Here $g=\hbar \sqrt{4 \pi a_{\rm bg} \Delta B \Delta \mu / m}$ is the 
atom-molecule coupling, $\Delta B$ is the width of the Feshbach resonance, 
$\Delta \mu$ is the difference in magnetic moments, 
and $a_{\rm bg}$ is the background scattering length.
\begin{figure}[Hb!]
\includegraphics[width=.9\columnwidth]{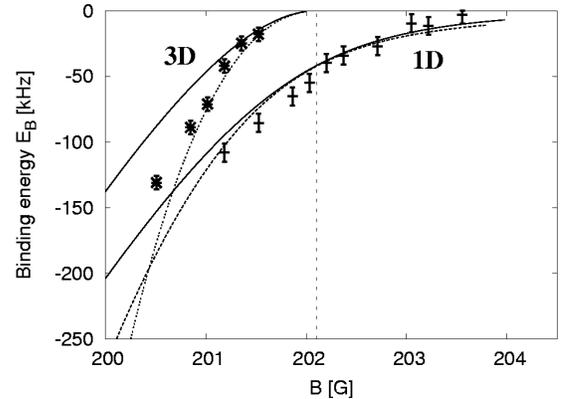}
\caption{
Binding energies for 1D and 3D molecules as a function of the magnetic field.
The solid lines correspond to the single-channel result. The dashed lines
are calculated within the two-channel theory.
}\label{fig1}
\end{figure}
In Fig. \ref{fig1} we
show for this three-dimensional case the molecular binding energy for both 
the single and two-channel approaches, respectively. Whereas the single-channel
results deviate significantly from the experimental data,
there is an excellent agreement with the two-channel theory.
It is therefore \emph{a priori} not clear that in the one-dimensional case 
the single-channel theory as given by Eq. (\ref{eq1}) is sufficiently 
accurate for the full range of magnetic fields explored by the experiment. 
In the following we derive the selfenergy for the confined case and make a comparisson with 
the experimental data.

\section{Theory}\label{secth}

Two atoms in a waveguide near a Feshbach resonance  
are described by the following hamiltonian,
\begin{eqnarray}
H = H_{\rm a} + H_{\rm m} + V_{\rm am}.
\end{eqnarray}
Here $H_{\rm a}$ represents the atomic contribution,
$H_{\rm m}$ describes the bare molecules, and $V_{\rm am}$ is the atom-molecule
coupling. Explicitely we have for the atoms,
\begin{eqnarray}
H_{\rm a} = \sum_{i=1,2} \left\{ K_{i} 
+ \frac{m \omega_{\perp}^{2}}{2} (x_{i}^{2} + y_{i}^{2}) \right\}
+ V_{\rm aa} \delta ( {\bf r}  ),
\end{eqnarray}
with $K_{i} = -\hbar^{2} \nabla^{2}_{i} / 2 m$ the kinetic energy of atom $i$,
$V_{\rm aa}$ is the strength of the nonresonant atom-atom interaction, and 
${\bf r}$ the relative coordinate of the two atoms.
The atoms are coupled to a molecular channel with a coupling $V_{\rm am}$. 
Near the resonance we have that $V_{\rm aa} \ll V_{\rm am}$, which allows us to 
neglect the nonresonant atom-atom interaction in that case.
For two atoms in the waveguide the two-channel Feshbach problem in the 
relative coordinate, after splitting off the center-of-mass motion, is then
given by,
\begin{eqnarray}\label{seq}
\left(
\begin{matrix}
H_{0}  & V_{\rm am} \\
V_{\rm am} & \delta_{B}
\end{matrix}
\right)
\left(
\begin{matrix}
|\psi_{\rm a} \rangle  \\
|\psi_{\rm m} \rangle
\end{matrix}
\right)
= E
\left(
\begin{matrix}
|\psi_{\rm a} \rangle  \\
|\psi_{\rm m} \rangle
\end{matrix}
\right).
\end{eqnarray}
Here the atomic Hamiltonian is 
$H_{0} = - \hbar^2 \nabla^2_{\bf r} / m + {\bf r}_{\perp}^{2}/4$, where
$\nabla^2_{\bf r} = \partial_{{\perp}}^{2}
+ \partial_{{z}}^{2}$ and ${\bf r}_{\perp}$ is the radial component of ${\bf r}$.
Only the relative part is relevant here, since only this part contains 
the interaction between the atoms.
The bare detuning is denoted by $\delta_{B}$.
The eigenstates  $|\psi_{n,k_{z}}\rangle$ of $H_{0}$ that are relevant for an s-wave Feshbach resonance 
are a product state of a two-dimensional harmonic oscillator wave function in the radial direction
and a plane wave along the axial direction.
The associated energies are given by
\mbox{$E_{n,k_{z}} = (2 n + 1) \hbar \omega_{\perp} + \hbar^{2} k_{z}^{2} / m$.}
The eigenstates of the two-dimensional harmonic oscillator that 
are relevant for s-wave scattering can be written
as $\psi_{n} (r_{\perp},\phi) = 
\left(2 \pi a_{\perp}^{2} \right)^{-1/2} e^{- r_{\perp}^{2}/4 a_{\perp}^{2}} 
~{L_{n}^{(0)}(r_{\perp}^{2} /2 a_{\perp}^{2})}$, where 
 $L_{n}^{(0)}(x)$ is the generalized Laguerre polynomial
and $\hbar \omega_{\perp} = \hbar^{2} / m a_{\perp}^{2}$. 
From Eq. (\ref{seq}) we obtain the following equation determining the binding energy of the molecules:
\begin{eqnarray}\label{seq2}
\langle \psi_{\rm m} | V_{\rm am} 
\frac{1}{E - H_{0}} V_{\rm am} | \psi_{\rm m} \rangle = E - \delta_{B}.
\end{eqnarray}
Using the above mentioned eigenstates of $H_{0}$,  Eq. (\ref{seq2}) can be written as
\begin{eqnarray}
\sum_{n=0}^{\infty} \int \frac{d k_{z}}{2 \pi}~
\frac{|\langle \psi_{\rm m} | V_{\rm am}| \psi_{n, k_{z}} \rangle  |^{2}}{E - E_{n,k_{z}}}  = E - \delta_{B}.
\end{eqnarray}
Using also the usual pseudopotential approximation for the atom-molecule
coupling, we have that 
$\langle {\bf r} | V_{\rm am} | \psi_{\rm m} \rangle = 
g \delta ({\bf r})$.
Substituting this and performing the $k_{z}$ integration we obtain
\begin{eqnarray}
E - \delta_{B} &=& \lim_{r_{\perp} \downarrow 0}
\frac{-  g^{2} m}{\sqrt{2} (4 \pi a_{\perp} \hbar^{2})}
 \nonumber \\ \times && \sum_{n=0}^{\infty} 
\frac{e^{-r_{\perp}^{2} /4 a_{\perp}^{2} } 
~{L_{n}^{(0)}( r_{\perp}^{2}/ 2 a_{\perp}^{2} ) }}{
\sqrt{n + 1/2 - E/ 2 \hbar \omega_{\perp} }}. 
\end{eqnarray}
The inverse square root 
$1/ \sqrt{n + 1/2 - E/ 2 \hbar \omega_{\perp} }$ in the summand
can be represented by the integral 
$ (2/\sqrt{\pi}) \int_{0}^{\infty} dt~ e^{- (n + 1/2 - E/ 2 \hbar \omega_{\perp})~t^{2}} $. 
To evaluate the sum over $n$ we substitute the above integral representation. 
The dependence on $n$ of the summand appears now in the exponent 
and in the degree of the Laguerre polynomial.
As a result the sum can be directly evaluated by 
making use of the generating functions of the Laguerre polynomials,
\begin{equation}
\sum_{n=0}^{\infty} L_{n}^{(0)}(x)~z^{n} = (1 - z)^{-1} 
\exp{\left( \frac{x z}{z-1} \right)}.
\end{equation}
In our case we have $z = e^{-t^{2}}$.
Using this result and making the transformation $y = t^{2}$ we arrive at
\begin{eqnarray}\label{11}
E - \delta_{B} &=&
\lim_{r_{\perp} \downarrow 0}
\frac{-  g^{2} m}{\sqrt{2 \pi} (4 \pi a_{\perp} \hbar^{2})}
 \int_{0}^{\infty} 
\exp{\left( \frac{r_{\perp}^{2} }{2 a_{\perp}^{2}}  \frac{e^{-y}}{e^{-y} - 1} \right)} 
\nonumber \\ \times &&
\frac{\exp{\left\{ -(1/2 - E/ 2 \hbar \omega_{\perp})~y \right\}}}{
\sqrt{y}~ (1 - e^{-y})} ~dy
\end{eqnarray}
For small values of $y$ the integrand in the above equation behaves as
$y^{-3/2} e^{-r^{2}/2 y}$. Note that we have
\begin{equation}
\frac{1}{\sqrt{\pi}} \int_{0}^{\infty} dy~ y^{-3/2}e^{r^{2}/2 y}
= \sqrt{2}/r .
\end{equation}
We add and subtract this integral from Eq. (\ref{11}) and in doing so
we explicitely  split off the $1/r$ divergence from the sum. 
The divergence in the selfenergy is energy independent and is related to 
the ultraviolet divergence that comes about because we have used pseudopotentials.
To deal with this divergence we 
have to use the renormalized detuning instead of the bare detuning. The former is defined as 
$\delta = \delta_{B} - \lim_{{r} \downarrow 0} m g^{2} / 4 \pi \hbar^{2} {r}$,
where $\delta = \Delta \mu (B - B_{0})$ is determined by the experimental 
value of the magnetic field $B_{0}$ at resonance
and  the magnetic moment difference $\Delta \mu = 16/9$ 
Bohr magneton for the ~$^{40}$K atoms of interest.
Note that, as expected, the required subtraction is exactly equal to the one needed
in the absence of the optical lattice.
In the latter case we have to subtract 
$g^{2}  \int d {\bf k}~m / \hbar^{2} {\bf k}^{2} (2 \pi)^{3}$ \cite{review,Holland},
which can be interpreted as $\delta = \delta_{\rm B} - \lim_{{\bf r} \downarrow 0} 
 g^{2} \int d {\bf k}~e^{i {\bf k}\cdot {\bf r} }m / \hbar^{2} {{\bf k}^{2}} (2 \pi)^{3}$.
Using the renormalized detuning we find that the binding energy of the dressed molecules 
satisfies the desired equation 
\begin{eqnarray}\label{eq2}
E_{B} - \delta(B) = \hbar \Sigma (E_{B}),
\end{eqnarray}
where the molecular selfenergy for the harmonically confined one-dimensional system 
is given by
\begin{eqnarray}
\hbar \Sigma (E) = -\frac{ m g^{2} }{ \sqrt{2} \left(4 \pi a_{\perp} \hbar^{2} \right)}~
\zeta (1/2,1/2 - E/2 \hbar \omega_{\perp}).
\end{eqnarray}

\section{Results and Discussion}\label{secres}

Using the selfenergy for the confined gas
we can now solve for the binding energy in Eq. (\ref{eq2}). 
The result is also shown in Fig. \ref{fig1}.
We find an improved description of the experiment, although 
the differences with the single-channel prediction are small near resonance
and only become large for larger detunings.
\begin{figure}[Hb!]
\includegraphics[angle=270,width=.8\columnwidth]{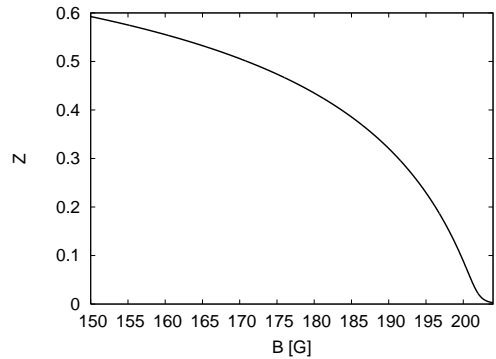}
\caption{
The bare molecule fraction $Z$ as a function of the magnetic field.
}\label{fig2}
\end{figure}
This presents one way in which to experimentally probe these differences.
Alternatively, it is also possible to directly measure the 
bare molecule fraction $Z$ of the dressed molecules \cite{Hulet}, which is always equal to zero in the 
single-channel model.
To be concrete we have for the dressed molecular wave function 
\begin{equation}
|\psi_{\rm dressed} \rangle = \sqrt{Z} |\psi_{\rm closed}\rangle + \sqrt{1 - Z} |\psi_{\rm open}\rangle,
\end{equation}
where $|\psi_{\rm closed} \rangle$ is the wave function of the bare molecules and $|\psi_{\rm open} \rangle$
denotes the wave function of the atom pair in the open channel of the Feshbach resonance. With this application in mind
we have plotted in Fig. \ref{fig2} also 
the probability $Z$, which is determined from the selfenergy by 
$Z = 1/(1 - \partial \hbar \Sigma (E_{B}) / \partial E_{B})$.

This work is supported by the Stichting voor Fundamenteel Onderzoek der
Materie (FOM) and by the Nederlandse Organisatie voor
Wetenschaplijk Onderzoek (NWO).

\bibliographystyle{apsrev}

\end{document}